\documentclass[11pt,twoside]{article}
\usepackage{asp2010}
\usepackage{natbib}

\aspvolume{471} 
\aspvoltitle{Origins of the Expanding Universe: 1912-1932}
\aspcpryear{2013} 
\aspvolauthor{Michael J. Way and Deidre Hunter, eds.} 

\begin{document}

\title{The Critical Importance of Russell's Diagram}

\author{Owen Gingerich
\affil{Harvard-Smithsonian Center for Astrophysics, Cambridge, MA, 02138, USA}}

\begin{abstract}
The idea of dwarf and giants stars, but not the nomenclature, was first 
established by Eijnar Hertzsprung in 1905; his first diagrams in support appeared in 1911.   
In 1913 Henry Norris Russell could demonstrate the effect far more strikingly because he 
measured the parallaxes of many stars at Cambridge, and could plot absolute magnitude
against spectral type for many points.  The general concept of dwarf and giant stars was
essential in the galactic structure work of Harlow Shapley, Russell's first graduate
student.  In order to calibrate the period-luminosity relation of Cepheid variables, he was
obliged to fall back on statistical parallax using only 11 Cepheids, a very sparse sample.
Here the insight provided by the Russell diagram became critical.  The presence of
yellow K giant stars in globular clusters credentialed his calibration of the
period-luminosity relation by showing that the calibrated luminosity of the Cepheids was
comparable to the luminosity of the K giants. It is well known that in 1920 Shapley did
not believe in the cosmological distances of Heber Curtis' spiral nebulae.  It is not so
well known that in 1920 Curtis' plot of the period-luminosity relation suggests that he
didn't believe it was a physical relation and also he failed to appreciate the significance
of the Russell diagram for understanding the large size of the Milky Way.
\end{abstract}

	Ninety years ago, when the Royal Astronomical Society was celebrating its centennial, 
the RAS President, A.S. Eddington selected six great landmarks of astronomical progress in that 
century.  He included the determination of stellar parallax, the discovery of Neptune, the rise of 
spectroscopic astronomy, and from the decade of relevance to our conference, the measurement 
of the angular diameter of Betelgeuse with Michelson's interferometer.  How astonishing to have 
this last item on the list, a measurement now so obscure that most of today's astronomy graduate 
students have never heard of it!  ``It seems to me," Eddington declared, ``to be worthy of a place 
in this select list as a triumph of scientific achievement which is second to
none."\footnote{\citet[][pp. 815-7]{Eddington1922Natur.109..815E}}

	While today we may be reluctant to include this measurement in such a list, it behooves 
us to parse its meaning a bit more to place its significance in the astronomical ferment of the 
teens of the 20th century, the time of Lowell, Slipher, Hale, Hertzsprung, Russell, Shapley, 
Curtis, Eddington, and Kapteyn, to name a few of the leaders.

	In 1897 in vol 28 of the Harvard College Observatory Annals, Antonia Maury and Annie 
J. Cannon had presented the first fruits of their spectral classification
work.  Annie Cannon's 
scheme was a refinement of the letter categories started earlier by Wilhelmina Fleming under 
E.~C. Pickering's general supervision\footnote{\cite{Cannon1901}}.  Cannon added the numbers, so that classes G0, G1, and 
G2 in principle followed F8 and F9, although not all the numbers were actually defined.  This 
was to become the basis of the Henry Draper Catalogue, issued in volumes 91 to 99 of the 
Harvard Annals (1918--24).
Maury's scheme had 22 categories designated with Roman 
numerals, each with three subdivisions, a, b, and c -- too complicated and clumsy for widespread 
use, most astronomers concluded.\footnote{\cite{Maury1897AnHar..28....1M}}
The Danish astronomer Eijnar Hertzsprung, then working in 
Leiden, thought otherwise and wrote to Pickering in 1908 that ``In my opinion the separation by 
Antonia Maury of the c and ac-stars is the most important advancement in stellar classification 
since the trials by Vogel and Secchi . . . .  To neglect the c-properties in classifying stellar 
spectra, I think, is nearly the same thing as if the zoologist, who has detected the deciding 
differences between a whale and a fish, would continue in classifying them
together."\footnote{E. Hertzsprung to E.~C. Pickering, 22 July 1908, Pickering
papers, Harvard University Archives.}

	In any event, Maury's classifications enabled Hertzsprung to make an interesting analysis 
in 1905.  What he showed was that the rare three dozen stars in Maury's subclasses c and ac, 
seemed intrinsically brighter than those in classes a and b. Since it would have
been difficult to get reliable parallaxes for the intrinsically brighter and
more distant stars, Hertzsprung drew his conclusion statistically using proper
motions.\footnote{An English translation by Harlow Shapley and Vincent Icke of
Hertzsprung's 1905 paper is printed in \citet[][pp. 209-211]{lang1979source}.}
This should have been seen as a 
peculiar and provocative result by the astronomical community, except that not many saw it.   
Hertzsprung had buried it and his follow-up paper of 1907 in a journal for photographic 
technology, \emph{Zeitschrift f\"{u}r wissenschaftliche
Photographie}.\footnote{\citet[][p. 86]{hertzsprung1905zeitschrift} and \citet[][p. 429]{hertzsprung1907zeitschrift}.}
Some years later, when Eddington 
found it, he chided Hertzsprung, saying, ``One of the sins of your youth -- to publish important 
papers in inaccessible places."\footnote{A.~S. Eddington to E. Hertzsprung,
7 August 1925, quoted in \citet[][p. 236]{nielsen1964}.}

	It must have taken a few decades for the physical reasons for the subclass c 
characteristics to be understood.  The normal diffuseness of the hydrogen lines in B, A, and F 
stars is caused by the Stark effect, where the electric fields generated by passing ions disturb the 
atomic excitation levels in the hydrogen atoms.  The effect is enhanced by the comparative 
density of main sequence stars.  Giant or supergiant stars have much lower particle densities in 
their atmospheres, so the atomic excitation levels are much less disturbed, and the spectral lines 
are distinctly sharper.  Note that Maury classified 649 brighter northern stars in her monograph, 
of which only 35 were labeled c or ac, all in spectral types that showed conspicuous hydrogen 
lines.

\begin{figure}[ht!]
\center{\includegraphics[scale=0.3,angle=-90]{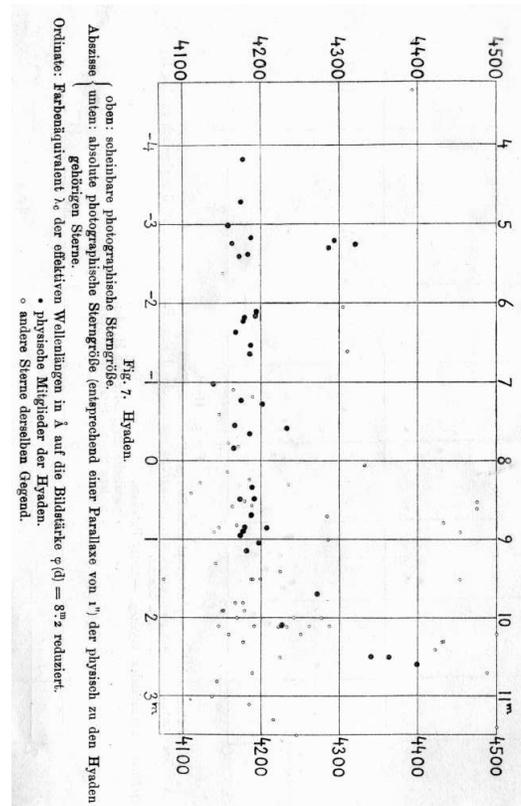}}
\caption{Hertzsprung's 1911 luminosity-color diagram for the Hyades, from the
Potsdam Astrophysical Observatory's \emph{Publikationen} 22, Nr. 63, p. 30.
The colors are specified by the effective
wavelength in {\AA}ngstroms, and the apparent magnitude is shown above. 
The diagram is here rotated 90$\deg$ for comparison with the Russell
diagram.}\label{gingerichfig01}
\end{figure}

	In a third paper, printed in 1911 in the Potsdam Astrophysical
Observatory's Publications\footnote{\cite{Hertzsprung1911POPot..63.....H}}
Hertzsprung actually included diagrams, plotting apparent magnitude against color for stars in 
the Pleiades and Hyades.   Because the stars in each cluster could be assumed to be at the same 
distance, this was equivalent to plotting the absolute magnitude against spectral type.  But the 
Pleiades cluster has no giants or supergiants, so its diagram showed no bifurcation.  In contrast, a 
few giants, scarcely even a handful, were present in the Hyades, once again showing a two-fold 
division of stars within their spectral types, but hardly dramatically so
(see Figure \ref{gingerichfig01}).

	Several years later, in 1913, Henry Norris Russell could demonstrate the
effect far more strikingly \footnote{\cite{Russell1913Obs....36..324R}} because
by that time he had access to the parallax for many stars, and thus he could
plot absolute magnitude against spectral type with a substantial number of
points. Russell distinguished the stars into dwarfs and giants, attributing the
nomenclature to Hertzsprung, who had not in fact used such a designation in
either of his \emph{Zeitschrift}
papers \footnote{\cite{hertzsprung1905zeitschrift,hertzsprung1907zeitschrift}}
or in the Potsdam publication.\footnote{\cite{Hertzsprung1911POPot..63.....H}}

	Many years ago I directly queried Hertzsprung about his possible
invention of the dwarf-giant nomenclature, and he replied: ``I hasten to
say that I have avoided the expressions `giant' and `dwarf', because the
stars are not very different in mass, but in density.  They are more or 
less `swollen'." Near the end of a popular lecture given at the end of 1908
(and published the following year in \emph{Himmel und
Erde})\footnote{\citet[][pp. 433-451]{schwartzschild1909}} Karl Schwarzschild used the
term giants (Giganten) repeatedly, but not dwarf.  So it is quite possible
that Russell himself invented the paired usage of both giant and dwarf.

\begin{figure}[ht!]
\center{\includegraphics[scale=0.6]{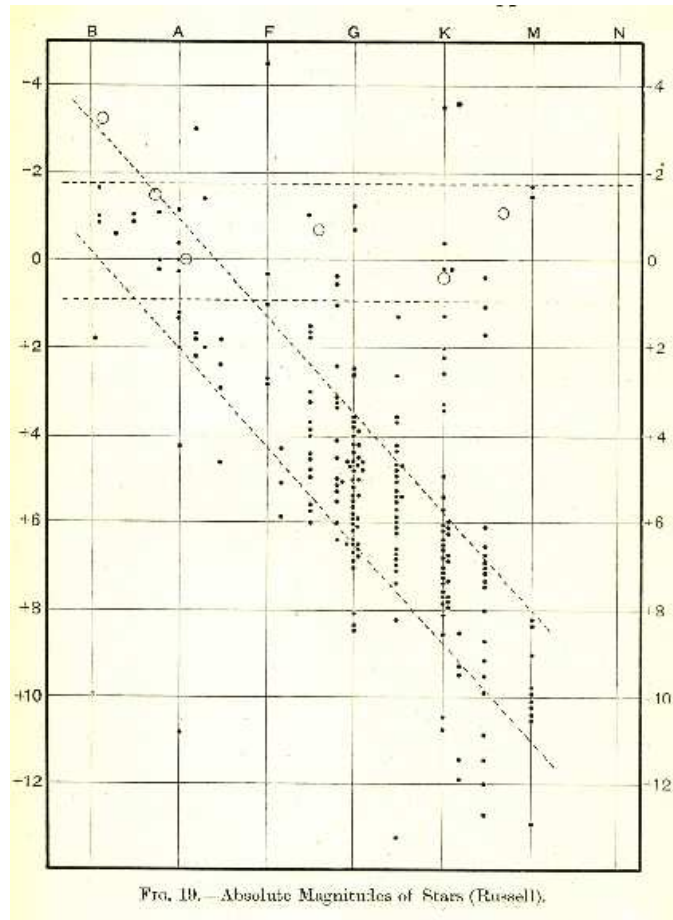}}
\caption{Russell's famous spectrum-luminosity diagram of 1913, as improved
typographically in Eddington's 1914 \emph{Stellar Movements and the
Structure of the Universe}.}\label{gingerichfig02}
\end{figure}

	This nomenclature was promptly picked up by \emph{Observatory} by
entitling Russell's report to the June 1913 meeting of the Royal Astronomical
Society simply ```Giant' and `Dwarf' Stars."\footnote{\cite{Russell1913Obs....36..324R}}
Russell and several other American astronomers had stopped in London on their 
way to the International Solar Union meeting in Bonn.  The RAS session gave Russell a much 
larger audience than at the Astronomy and Astrophysics Society of America meeting in Atlanta 
at the end of the year where what might be termed the official introduction of the Russell 
diagram took place.  Subsequently he published his paper with the diagram in
\emph{Popular Astronomy} \footnote{\cite{Russell1914PA.....22..275R}}
and in \emph{Nature} \footnote{\cite{Russell1914Natur..93..227R,Russell1914Natur..93..252R,Russell1914Natur..93..281R}}
-- the earlier \emph{Observatory} report noted that there had been slides, but 
didn't reproduce them (see Figure \ref{gingerichfig02}).

	For the moment it is enough to notice that much of astronomy was in flux
in this decade, from the source of stellar energy and its relevance to stellar
evolution, to the place of the spiral nebulae and structure of the Milky Way.
An unexpected discovery by J. C. Kapteyn, of star streaming, announced in 1904
at the St. Louis World's Fair,\footnote{Kapteyn's report, "Statistical Methods
in Stellar Astronomy" is published in \emph{International Congress of Arts and
Sciences, St. Louis}, 4, 396-425.  The following year he attended the British
Association for the Advancement of Science meeting in Capetown,
and then published his principal paper, "Star Streaming," in Report
of the British Association for the Advancement of Science, section A, 257-265.}
indicated that the motion of stars in the Milky Way was not random.
The physical basis for this effect would not be understood until 
Shapley's model of the Milky Way provided a basis for knowledge of
galactic rotation.  But star streaming would provide a central focus
of Eddington's first book, \emph{Stellar Movements and the 
Structure of the Universe},
published in 1914\footnote{\citep{Eddington1914QB801.E48}},
and in this book was one of the earliest presentations of Russell's diagram.
Russell had displayed his diagram to a crowded meeting the 
Royal Astronomical Society in June of 1913, and soon thereafter Eddington had
asked Russell for a copy.  It played an important role in the chapter where
Eddington related stellar movements to spectral classes.  

	In his book Eddington wrestled at length with the observation that early-type stars had 
distinctly smaller space velocities than K and M stars, and this was after removing effects of 
solar motion.  Ultimately it became an issue of luminosities.  In the parallax work, the M stars 
were the least luminous, whereas in the radial velocity work, the stars were systematically 
selected by magnitude since the brighter stars were more congenial to spectrographic analysis, 
and this picked up more highly luminous stars, stars that we now know to be giant stars.  ``The 
leading contribution to this problem," Eddington wrote as the book was in its final stages of 
preparation, ``is the hypothesis of `giant' and `dwarf' stars put forward by Hertzsprung and 
Russell.  They consider each spectral type to have two divisions, which are not in reality closely 
related."  And there in his book, across from that paragraph, is a typographically very clean 
rendition of the diagram that Russell had sent him.  This was the beginning of the resolution of 
Eddington's stellar motions paradox, but it would require both the knowledge of galactic rotation 
and of stellar populations before the puzzle would really be solved.  

\begin{figure}[ht!]
\center{\includegraphics[scale=1.0]{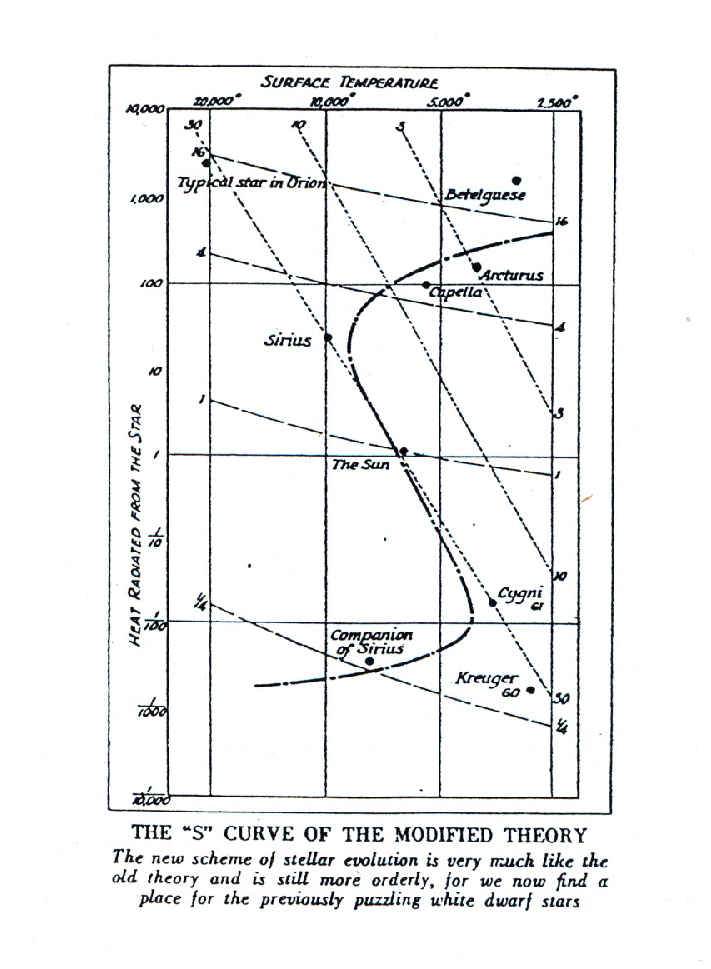}}
\caption{Russell's 1925 luminosity--surface temperature diagram with an
evolutionary track including the white dwarf companion of Sirius.}
\label{gingerichfig03}
\end{figure}

	For Russell the diagram provided a convenient way to illustrate his
ideas about stellar evolution.  In introducing his diagram, he noted that the
stellar densities increased from the giant branch down the dwarf sequence,
that is, what we call the main sequence.   ``If that is also the order of
advancing age, we are led at once back to Norman Lockyer's hypothesis that a
star is hottest near the middle of its history, and that the redder stars fall
into two groups, one of rising and the other of falling temperature," he wrote.
``The giant stars then represent successive stages in the heating up of a body,
and must be more primitive the redder they are; the dwarf stars represent
successive stages in later cooling, and the reddest of these are the farthest
advanced." Eventually Eddington's mass-luminosity relation would add evidence
for this evolutionary march down the main sequence, but that would come a decade
later.  When Eddington demonstrated that the main sequence was essentially a
mass sequence it seemed to follow that the stars had to move down the main
sequence as they gradually converted their active mass into energy.
Figure \ref{gingerichfig03} comes from Russell's 1925 paper
\footnote{\cite{Russell1925Natur.116..209R}}, and for the
first time includes a white dwarf in the evolutionary 
track.  I think Russell could hardly have imagined then what a powerful
evolutionary tool his diagram would eventually become in the hands of Walter
Baade and Martin Schwarzschild and their students including Sandage,
Arp and Baum.  

	The issue of the naming of the diagram smoldered for years.
This visual tool was for three decades generally referred to as
``the Russell diagram," although in 1933 Bengt Str\"{o}mgren 
introduced the designation Hertzsprung-Russell diagram, abbreviated
H.-R. in a paper in the \emph{Zeitschrift f\"{u}r
Astrophysik},\footnote{\cite{Stromgren1933ZA......7..222S}} although in his
famous paper there on the hydrogen content of stars he had in the previous year
used ``Russell diagram."\footnote{\cite{Stromgren1932ZA......4..118S}}
In the 1940s it was renamed the H-R diagram on the style sheet of
the \emph{Astrophysical Journal}, this done by co-editor Subrahmanyan 
Chandrasekhar to satisfy the relentless hectoring from fellow Chicago
faculty member, Gerard Kuiper.\footnote{\citet[][p. 329]{devorkin2000russell}}
Chandra, however, had already adopted Str\"{o}mgren's designations with a
chapter headed ``Str\"{o}mgren's Interpretation of the Hertzsprung-Russell
Diagram" and the use of the abbreviation H.R. within the chapter itself in his
1939 book \emph{Stellar Structure}.\footnote{\cite{Chandra1939isss.book.....C}}

\begin{figure}[ht!]
\center{\includegraphics[scale=0.5]{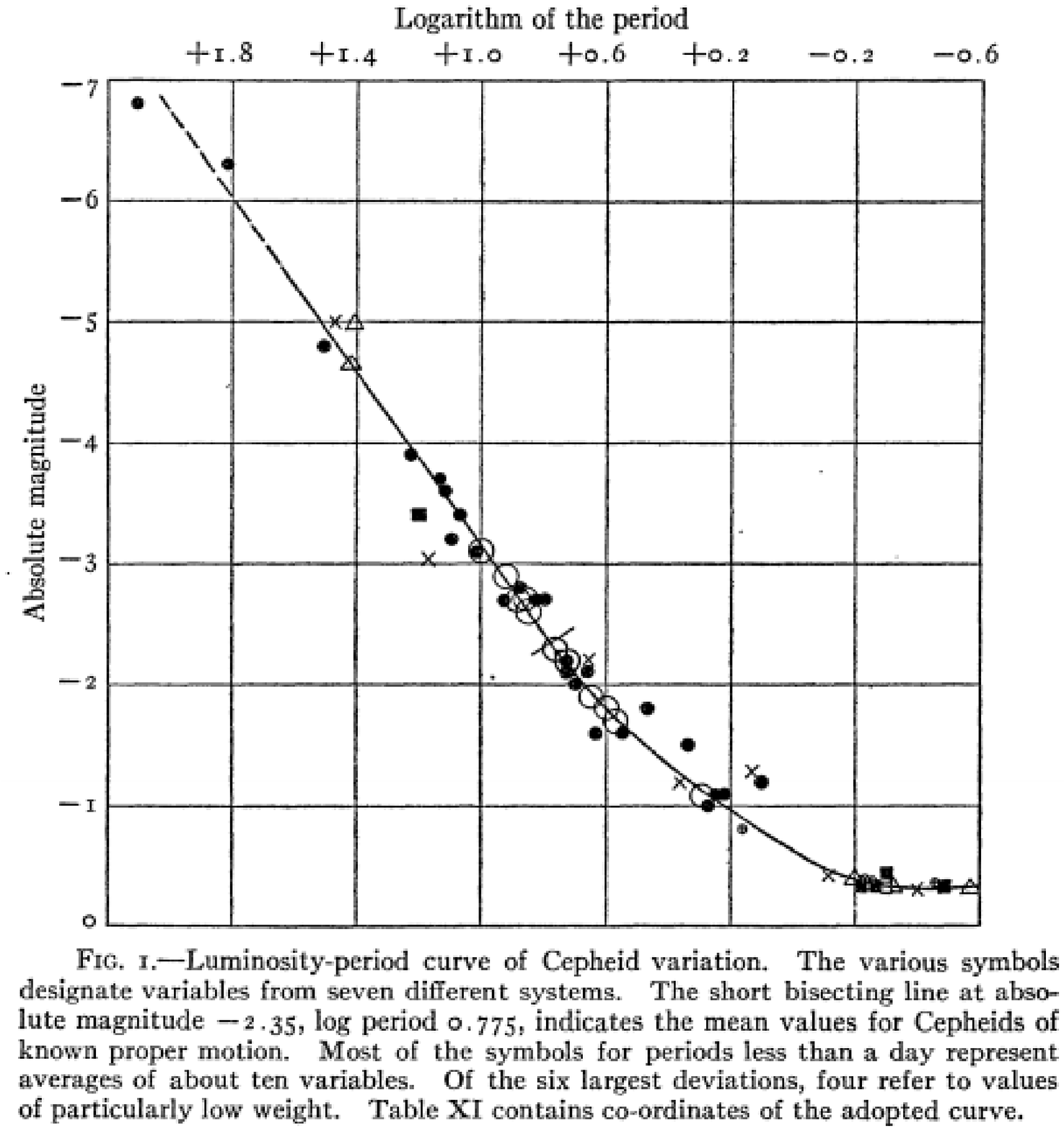}}
\caption{Shapley's well-honed period-luminosity curve for selected Cepheid variable stars,
from his ``Sixth Paper" in the 1918 \emph{Astrophysical Journal}.}\label{gingerichfig04}
\end{figure}

	But meanwhile, back in the nineteen-teens, the general concept of dwarf
and giant stars was essential in the galactic structure work of Harlow Shapley,
Russell's thesis student, now transplanted to Mount Wilson Observatory.
His research at Princeton had included observations and analysis of 90
eclipsing binaries, essentially raising the number of binary star orbits by an 
order of magnitude.  Now, with the 60-inch reflector, he could attack the problem
of variable stars in globular clusters, and this led to the calibration of
Henrietta Leavitt's period-luminosity relation.  Since none of the field
Cepheids in the Milky Way were close enough for trigonometric parallaxes,
Shapley was obliged to fall back on statistical parallax using the proper
motions of 11 Cepheids, which is a pretty sparse sample.  In a paper written
in November, 1917, Shapley discussed his calibrated
period-luminosity diagram (see Figure \ref{gingerichfig04}),
beautifully and unbelievably smoothed by averaging triplets of the basic
set of eleven Cepheids and thus adjusting their absolute
magnitudes.\footnote{\cite{Shapley1918ApJ....48...89S}}
Earlier, in his graduate studies at Princeton, he had demonstrated that
the common assumption that the Cepheids were spectroscopic binary stars
led to the ridiculous conclusion that one star was inside
the other!\footnote{\cite{Shapley1914ApJ....40..448S}}

His counter-suggestion, a pulsation hypothesis, meant that the Cepheids were
physical objects.  Because Shapley firmly believed that a physical law
connected the pulsation periods of these stars and luminosities, he felt
his smoothing procedure was fully justified.    

Now the insight provided by the Russell diagram, including the giant-dwarf
distinction, became critical.  The brightest stars in the globular
clusters included Cepheid variables and yellow K stars.  Shapley recognized
that the yellow K stars were giant stars, whose absolute 
magnitudes were approximately known from giant galactic field stars.
This credentialed his calibration of the period luminosity relation by
showing that the calibrated luminosity of the Cepheids was comparable to the
luminosity of the K giants.  Without referring specifically to the Russell
diagram, he nevertheless used the understanding it provided for this comparison
in another paper of his 1917 series (but which was not published until 1919
because he was overwhelming the \emph{Astrophysical Journal} with research
results).\footnote{\cite{Shapley1919ApJ....49...24S}}

\begin{figure}[ht!]
\center{\includegraphics[scale=0.6]{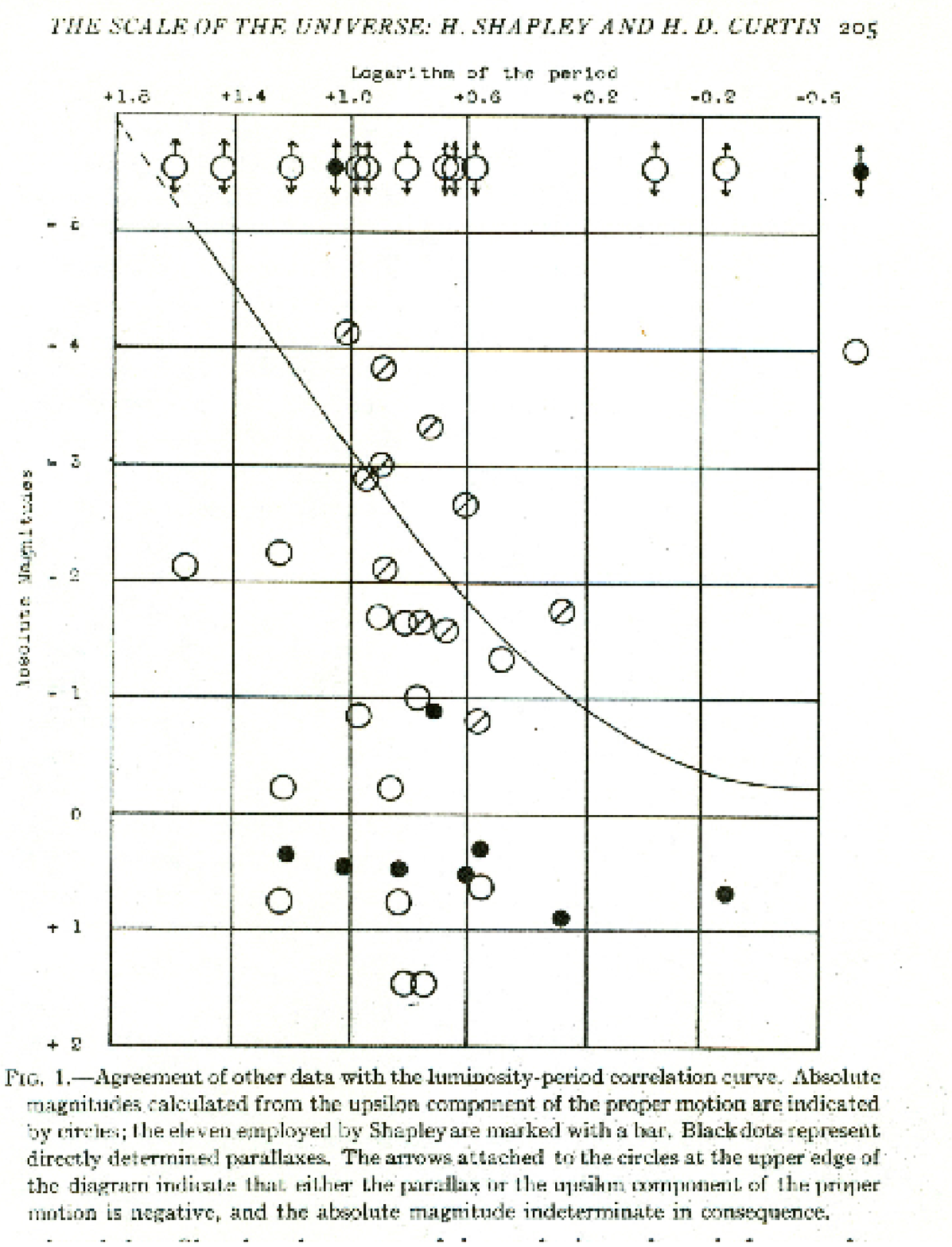}}
\caption{H. D. Curtis' period-luminosity curve for Cepheid variable stars,  from the
Shapley-Curtis debate as published in the \emph{Bulletin of the National Academy of Sciences}, May, 1921.}
\label{gingerichfig05}
\end{figure}

In the celebrated Shapley-Curtis debate sponsored by the National Academy of
Sciences in Washington in April, 1920, Heber D. Curtis blasted Shapley's
period-luminosity diagram for the Cepheid variables.
That the results of Shapley's avant-garde statistical parallax technique had
not convinced him is clearly shown in his own version of the putative
relationship, which looks like the target of a beginner marksman,
essentially a scatter diagram
(see Figure \ref{gingerichfig05}).\footnote{\citet[][pp. 194-217]{Curtisnational1921bulletin}}
His diagram suggests he didn't believe in the period-luminosity relation!

	In the revised post-debate version of his 1920 encounter with Curtis,
Shapley pushed back, stating that his estimates of the distance scale
in the Milky Way did not depend on the calibration of the Cepheid variable
stars, but could be established through the giant
stars.\footnote{\citet[][pp. 171-193]{Shapleynational1921bulletin}}
Of course the understanding of the Russell diagram and the concept of giant
and dwarf stars was essential.  Shapley wrote:

\begin{quote}
 	An argument much insisted upon by Curtis is that the average absolute
magnitude of stars around the sun is equal to or fainter than solar brightness,
hence, that average stars we see in clusters are also dwarfs.  Or, put in a
different way, he argues that since the mean spectral class of a globular cluster
is of solar type and the average solar-type star near the sun is of solar
luminosity, the stars photographed in globular clusters must be of solar 
luminosity, hence not distant.  This deduction, he holds, is in compliance with
[the] proposition [of] uniformity throughout the universe.  But in drawing the
conclusions, Curtis apparently ignores, first, the very common existence of
red and yellow giant stars in stellar systems. And second . . . in treating a
distant system we naturally first observe its giant stars. . . .

	Suppose that an observer, confined to a small valley, attempts to
measure the distances of surrounding mountain peaks.  Because of the short base
line allowed him, his trigonometric parallaxes are valueless except for the
nearby hills.  On the remote peaks, however, his telescope shows green foliage.
First, he assumes approximate botanical uniformity throughout all visible
territory.  Then he finds the average height of all plants 
immediately around him (conifers, palms, asters, clovers, etc.) is one foot.
Correlating this average height of all plants visible against the sky line on
the distant peaks he obtains values of the distances. If, however, he had used
some method of distinguishing various floral types, he would not have mistaken
pines for asters and obtained erroneous results 
for the distances of the surrounding mountains.  
\end{quote}

	This critical understanding, then, depended on the efficacy of Russell's
dwarf and giant distinction, which was about to be verified by direct measurement
of the giant size of Betelgeuse with a Michelson interferometer attached to
the 100-inch reflector at Mt. Wilson.  It is well known that in 1920 Shapley no longer
believed in the cosmological distances of Curtis' spiral nebulae, which were
based largely on novae.  It is not so well known that in 1920 Curtis failed to
appreciate the significance of the Russell diagram for understanding 
the large size of the Milky Way, and the key role of the period-luminosity
relation.

	In 1924, after Hubble found his first Cepheid in M31, it was
to Shapley that he wrote for the calibrated period-luminosity
curve.\footnote{E. Hubble to H. Shapley, 19 February 1924, Shapley papers,
Harvard University Archives.}

	Within just a few years both Curtis and Shapley fundamentally changed
their views concerning the size of the Milky Way and the nature of the spirals.
In 1966 I took my English colleague Michael Hoskin for an oral history interview
with Shapley, and we found that the 81-year-old astronomer could scarcely even
imagine a time when he thought spirals were not distant galaxies.  Tempus fugit!
Beware of oral history interviews!

\nocite{Hub-Shap-24}
\nocite{Hertz-Pick-08}
\nocite{Hertzsprung1911POPot..63.....H}
\bibliography{gingerich}

\end{document}